\documentclass[iop,numberedappendix]{emulateapj-rtx4}
\usepackage{graphicx,natbib,color,bm,url,times}
\graphicspath{{./fig/}{./png/}}

\newcommand{\EQ}{\begin{equation}}
\newcommand{\EN}{\end{equation}}
\newcommand{\EQA}{\begin{eqnarray}}
\newcommand{\ENA}{\end{eqnarray}}

\newcommand{\Eq}[1]{Equation~(\ref{#1})}

\newcommand{\Sec}[1]{Section~\ref{#1}}

\newcommand{\Fig}[1]{Figure~\ref{#1}}

\newcommand{\Figp}[2]{Figure~\ref{#1}({#2})}

\newcommand{\Tab}[1]{Table~\ref{#1}}

\newcommand{\bra}[1]{\langle #1\rangle}

{}
{}
{}

{}
{}
{}
{}
{}
{}
{}
{}
{}
{}
{}
{}
{}
{}
{}
{}
{}

{}

\newcommand{\meanB}{\overline{B}}
\newcommand{\meanE}{\overline{E}}

\newcommand{\hatkk}{\hat{\bm{k}}}

{}

{}
{}
{}

\newcommand{\kk}{\bm{k}}
\newcommand{\KK}{\bm{K}}

\newcommand{\aaaa}{\bm{a}}
\newcommand{\bb}{\bm{b}}
\newcommand{\jj}{\bm{j}}

\newcommand{\grav}{\bm{g}}

\newcommand{\FF}{\bm{F}}

\newcommand{\uu}{\bm{u}}

\newcommand{\ee}{\mbox{\boldmath $e$} {}}

\newcommand{\nab}{{\bm{\nabla}}}
\newcommand{\GG}{\mbox{\boldmath $G$} {}}

\newcommand{\OO}{\bm{\Omega}}

\newcommand{\SSSS}{\mbox{\boldmath ${\sf S}$} {}}

\newcommand{\ii}{{\rm i}}

\newcommand{\SKEW}{{\rm skew}}
\newcommand{\KURT}{{\rm kurt}}

\newcommand{\DD}{{\rm D} {}}

\newcommand{\dd}{{\rm d} {}}

\newcommand{\const}{{\rm const}  {}}

\def\la{\mathrel{\mathchoice {\vcenter{\offinterlineskip\halign{\hfil
$\displaystyle##$\hfil\cr<\cr\sim\cr}}}
{\vcenter{\offinterlineskip\halign{\hfil$\textstyle##$\hfil\cr<\cr\sim\cr}}}
{\vcenter{\offinterlineskip\halign{\hfil$\scriptstyle##$\hfil\cr<\cr\sim\cr}}}
{\vcenter{\offinterlineskip\halign{\hfil$\scriptscriptstyle##$\hfil\cr<\cr\sim\cr}}}}}

\def\Pm{\mbox{\rm Pr}_M}

\def\Imag{\mbox{\rm Im}}

\def\cp{c_{\rm p}}

\def\urms{u_{\rm rms}}

\def\Beq{B_{\rm eq}}

\def\half{{\textstyle{1\over2}}}

\newcommand{\nm}{\,{\rm nm}}

\newcommand{\Mm}{\,{\rm Mm}}

\newcommand{\yapj}[3]{ #1, {ApJ,} {#2}, #3}

\newcommand{\yapjl}[3]{ #1, {ApJ,} {#2}, #3}

\newcommand{\yan}[3]{ #1, {Astron.\ Nachr.,} {#2}, #3}

\newcommand{\yana}[3]{ #1, {A\&A,} {#2}, #3}

\newcommand{\yaraa}[3]{ #1, {ARA\&A,} {#2}, #3}

\newcommand{\yprl}[3]{ #1, {Phys.\ Rev.\ Lett.,} {#2}, #3}

\newcommand{\ymn}[3]{ #1, {MNRAS,} {#2}, #3}

\newcommand{\ysph}[3]{ #1, {Solar Phys.,} {#2}, #3}

\newcommand{\yprd}[3]{ #1, {Phys.\ Rev.\ D,} {#2}, #3}
\newcommand{\ypre}[3]{ #1, {Phys.\ Rev.\ E,} {#2}, #3}

\newcommand{\yjour}[4]{ #1, {#2}, {#3}, #4}

\newcommand{\ybook}[3]{ #1, {#2} (#3)}

\newcommand{\pana}[2]{ #1, {A\&A}, in press, arXiv:#2}
\newcommand{\dana}[3]{ #1, {A\&A}, DOI:#2, arXiv:#3}
\newcommand{\papj}[2]{ #1, {ApJ}, in press, arXiv:#2}

\newcommand{\arxiv}[2]{ #1, arXiv:#2}
\begin{document}
\title{$E$ and $B$ polarizations from inhomogeneous and solar surface turbulence}
\author{Axel Brandenburg$^{1,2,3,4,5}$},
\author{Andrea Bracco$^3$}
\author{Tina Kahniashvili$^{5,6}$}
\author{Sayan Mandal$^{5}$}
\author{Alberto Roper Pol$^{1,7}$}
\author{Gordon J. D. Petrie$^8$}
\author{Nishant K. Singh$^{9}$}

\affil{
$^1$Laboratory for Atmospheric and Space Physics, University of Colorado, Boulder, CO 80303, USA\\
$^2$JILA and Department of Astrophysical and Planetary Sciences, University of Colorado, Boulder, CO 80303, USA\\
$^3$Nordita, KTH Royal Institute of Technology and Stockholm University, Roslagstullsbacken 23, SE-10691 Stockholm, Sweden\\
$^4$Department of Astronomy, AlbaNova University Center, Stockholm University, SE-10691 Stockholm, Sweden\\
$^5$McWilliams Center for Cosmology and Department of Physics,
  Carnegie Mellon University, 5000 Forbes Ave, Pittsburgh, PA 15213, USA\\
$^6$Abastumani Astrophysical Observatory, Ilia State University,
  3-5 Cholokashvili St., 0194 Tbilisi, Georgia\\
$^7$Department of Aerospace Engineering Sciences, University of Colorado,
  Boulder, CO 80303, USA\\
$^8$National Solar Observatory, 3665 Discovery Drive, Boulder, CO 80303, USA\\
$^9$Max-Planck-Institut f\"ur Sonnensystemforschung, Justus-von-Liebig-Weg 3, D-37077 G\"ottingen, Germany
}
\submitted{\today,~ $ $Revision: 1.104 $ $}

\begin{abstract}
Gradient- and curl-type or $E$- and $B$-type polarizations have been
routinely analyzed to study the physics contributing to the
cosmic microwave background polarization
and galactic foregrounds.
They characterize the parity-even and parity-odd properties of the
underlying physical mechanisms, for example hydromagnetic
turbulence in the case of dust polarization.
Here we study spectral correlation functions characterizing the
parity-even and parity-odd parts of linear polarization for homogeneous
and inhomogeneous turbulence to show that only the inhomogeneous helical
case can give rise to a parity-odd polarization signal.
We also study nonhelical turbulence and suggest that a strong
nonvanishing (here negative) skewness of the $E$ polarization is
responsible for an enhanced ratio of the $EE$ to the $BB$
(quadratic) correlation in both helical and nonhelical cases.
This could explain the enhanced $EE/BB$ ratio observed recently for
dust polarization.
We close with a preliminary assessment of using linear polarization of
the Sun to characterize its helical turbulence without being subjected
to the $\pi$ ambiguity that magnetic inversion techniques have to address.
\end{abstract}

\email{brandenb@nordita.org}

\section{Introduction}

Helicity characterizes the swirl of a flow or a magnetic field.
Examples include the cyclonic and anticyclonic flows on a weathermap,
which are systematically different in the northern and southern hemispheres.
Similar differences are also seen on the solar surface, where both
flow and magnetic field vectors show swirl.
Both fields play important roles in the solar dynamo, which is believed to be
responsible for the generation of the Sun's large-scale magnetic field
\citep{Mof78,KR80,BS05}.

To determine the solar magnetic helicity, one first needs to determine
the magnetic field $\bb$.
This is done by measuring all four Stokes parameters to compute $\bb$
at the visible surface.
Historically, the first evidence for a helical magnetic field came from
estimates of the mean current helicity density, $\bra{\jj\cdot\bb}$,
where $\jj=\nab\times\bb/\mu_0$ is the current density and $\mu_0$
is the vacuum permeability.
Under the assumption of isotropy, using local Cartesian coordinates,
we have $\bra{\jj\cdot\bb}\approx\bra{j_z b_z}/3$, where
$j_z=(\partial_x B_y-\partial_y B_x)/\mu_0$ is the vertical component of
the current density, which involves only horizontal derivatives.
Another approach is to assume that the magnetic field above the solar
surface is nearly force-free.
A vanishing Lorentz force ($\jj\times\bb=\bm{0}$) implies that $\jj$ is
parallel to $\bb$, so $j_z=\alpha b_z/\mu_0$ with some coefficient $\alpha$.
The sign of $\alpha$ is directly related to the sign of the local current
helicity density.
\cite{See90} and \cite{PCM95} computed $\alpha$ and found it to be
negative in the northern hemisphere and positive in the southern.
This is a statistical result that has been confirmed many times
since then; see e.g., \cite{Singh18}.

A general difficulty in determining magnetic helicity lies in the fact
that the horizontal magnetic field components are only determined up to
the $\pi$ ambiguity.
In other words, one only measures horizontal magnetic field vectors
without arrow heads.
The actual horizontal field direction is usually ``reconstructed''
by comparing with that expected from a potential magnetic field
extrapolation, which only depends on the vertical magnetic field.
It is unclear to what extent this assumption introduces errors and how
those affect, for example, the scale dependence of the magnetic helicity
that was determined in some of the aforementioned approaches \citep[see,
e.g.,][]{ZBS14,ZBS16,BPS17,Singh18}.

To address the question about the limitations resulting from the $\pi$
ambiguity, we study here another potential proxy of magnetic helicity
that is independent of the $\pi$ ambiguity.
To this end, we decompose Stokes $Q$ and $U$, which characterize linear
polarization, into the $E$ and $B$ polarizations that are routinely
used in the cosmological context \citep{Kamion97,SZ97} as polarized
emission from the cosmic microwave background (CMB).
The $E$ and $B$ polarizations characterize parity-even and parity-odd
contributions; see \cite{KK99} for a review.
In the CMB, correlations of $B$ polarization with parity-even quantities
such as intensity, temperature, or the $E$ polarization
are believed to be proxies of the helicity of the underlying magnetic
field \citep{KMLK14}.

Observations of $E$ and $B\,$ polarizations have been obtained at various
frequencies using the {\em Planck} satellite.
Much of the $B$ polarization is now believed to come from the galactic
foregrounds including gravitational lensing, for example.
While a definitive $EB$ cross correlation has not yet been detected,
we do know that the $EE$ correlation is about twice as large as the $BB$
correlation in the diffuse emission \citep{Adam16}.
This was unexpected at the time \citep{CHK17} and will also be addressed in
this paper.

\cite{Kritsuk18} have shown that supersonic hydromagnetic turbulent
star formation simulations are able to reproduce the observed $EE/BB$ ratio
of about two, but the physical reason for this was not identified.
\cite{KLP17} discussed the dominance of magnetic over kinetic energy
density as an important contributor for causing the enhanced ratio.
Here, instead, we identify a strong skewness of the intrinsic $E$
polarization at all depths along the line of sight as an important
factor.

Our main objective is to study the connection between $EB$ cross correlation
and magnetic or kinetic helicities in various turbulent flows, which
can be related to what happens at the surface of the Sun.
As we will show in this paper, such a connection exists only
under certain inhomogeneous conditions, for example in the proximity of
a surface above the solar convection zone.
This automatically gives preference to one over the other viewing
direction of the plane perpendicular to the line of sight.
In homogeneous turbulence, by contrast, there is no way of differentiating
one viewing direction from the other.

\begin{figure}[t!]\begin{center}
\includegraphics[width=\columnwidth]{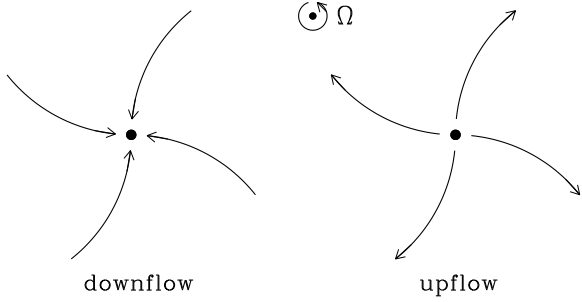}
\end{center}\caption[]{
Sketch illustrating the same shape of streamlines
for updrafts and downdrafts for convection in the northern
hemisphere ($\grav\cdot\OO<0$), as viewed from top down.
}\label{sketch}\end{figure}

Certain physical circumstances may well give preference to one over the
other side of a plane.
The CMB may indeed be one such example.
Rotating stratified convection is another rather intuitive example, as
illustrated in \Fig{sketch}.
In this sketch one sees the flow lines of cyclonic convection around
down- and upflows in the northern hemisphere as viewed from the top, so
all converging inflows attain a counterclockwise swirl, and all
diverging outflows attain a clockwise swirl.
As seen from the sketch, however, the orientations of both flow patterns
consisting of unsigned flow lines is the same.
This curl-type pattern gives rise to $B$-type
polarization of positive sign \citep[e.g.][]{KMLK14},
as will be verified in the next section.
It is only if we were to flip this plane or
when inspecting this pattern from beneath that we will
see a mirror image of the original pattern and therefore the opposite
sign of the $B$ polarization; see \cite{Dur08}.
Here, the $E$ polarization is the same when viewed from beneath (or in
a mirror).
This gives rise to a systematic $EB$ correlation.
The $E$ polarization will also be positive in this case if a ring-like
pattern dominates over a star-like pattern.
This results in a positive $EB$ correlation
in the north and negative in the south.

The purpose of this work is to use various numerical simulations
to determine the relation between magnetic helicity and the $EB$
correlation that is derived from just the horizontal field vectors without
knowledge of which of the two horizontal directions the vector points into
(i.e., the $\pi$ ambiguity).
The simulations include decaying homogeneous helical and nonhelical
turbulence and rotating stratified convection, which can serve as a
prototype for convective turbulence at the solar surface.

\section{$E$ and $B$ polarization}

\subsection{Formalism}

We consider magnetic field vectors, $\bb=(b_x, b_y, b_z)$, in one or
several two-dimensional $xy$ cross-sections in a three-dimensional volume.
We assume that this magnetic field affects the polarization of the
electromagnetic radiation whose electric field vectors in the
$(x,y)$ plane are $\ee=(e_x,e_y)$.
It is convenient to use complex notation and write this vector as
$e_x+\ii e_y\equiv|e|\,\exp(\ii\psi_e)$, where $\psi_e$ is the angle
of the electric field with the $x$ axis.
Likewise, we write the magnetic field in the plane, $(b_x,b_y)$,
as $b_x+\ii b_y\equiv|b|\,\exp(\ii\psi_b)$.
For electromagnetic radiation, electric and magnetic field vectors are
at right angles to each other, so $\psi_e=\psi_b+\pi/2$.
In complex form, the intrinsic (or local) polarization $p(x,y,z)$ is
proportional to the square of the complex electric field,
i.e., $p\propto(e_x+\ii e_y)^2\propto-\epsilon\,(b_x+\ii b_y)^2$,
where $\epsilon(x,y,z)$ is the local emissivity.
The magnetic field of the electromagnetic radiation aligns with the
ambient magnetic field, so that
\EQ
p=-\epsilon\,(b_x+\ii b_y)^2/\bb^2.
\EN
In most of the cases, we assume $\epsilon\propto\bb^2$,
which would be appropriate for the Sun \citep{SL87,BDTSMW14}.
For dust polarization, on the other hand, we assume
$\epsilon$ to be independent of $|\bb|$; see
\cite{PlanckXX} and \cite{Bracco18} for details.
The observable complex polarization, $P=Q+\ii U$,
is the line-of-sight integral
\EQ
P(x,y)=\int p(x,y,z)\,e^{-\tau(x,y,z)}\,\dd z,
\label{LOSintegr}
\EN
with $\tau(x,y,z)$ being the optical depth with respect to the observer.
If the medium can be considered optically thin, as for diffuse dust emission
in the interstellar medium, we can set $\tau=0$.
This will also be done in the present work.
In addition, we study the polarization from individual slices,
which corresponds to an optically thick case for that slice.

Next, we define $R=E+\ii B$, where $E$ and $B$ are the parity even and
parity odd contributions to the complex polarization, respectively.
They are related to each other in Fourier space via \citep{SZ97,Kamion97,Kamion97b,KK16}
\begin{equation}
\tilde{R}(k_x,k_y)=(\hat{k}_x-\ii \hat{k}_y)^2\tilde{P}(k_x,k_y),
\end{equation}
where $\hat{k}_x$ and $\hat{k}_y$ are the $x$ and $y$ components
of the planar unit vector $\hatkk=\kk/k$, with $\kk=(k_x,k_y)$ and
$k=(k_x^2+k_y^2)^{1/2}$ being the length of $\kk$.
Tildes indicate Fourier transformation over $x$ and $y$.
We transform $\tilde{R}$ back into real space to obtain $E(x,y)$ and $B(x,y)$
at a given position $z$.
We plot contours of $E$ and $B$ and overplot polarization vectors with angles
\EQ
\chi_E=\half\arg P_E \quad\mbox{and}\quad
\chi_B=\half\arg P_B,
\EN
where $P_E$ and $P_B$ are computed in Fourier space as
$\hat{P}_E=(\hat{k}_x+\ii \hat{k}_y)^2\tilde{E}$ and
$\hat{P}_B=(\hat{k}_x+\ii \hat{k}_y)^2(\ii\tilde{B})$.

We consider shell-integrated spectra along a ring of radius $k=|\kk|$
in wavenumber space of the form
\begin{equation}
C_{XY}(k)=\int_0^{2\pi}\tilde{X}(\kk)\,\tilde{Y}^*(\kk)\,k\,\dd\phi_k,
\label{KKnull}
\end{equation}
where $\phi_k$ is the azimuthal angle in Fourier space, and
$\tilde{X}$ and $\tilde{Y}$ are Fourier transformed fields
that each (or both) stand for $\tilde{E}$ or $\tilde{B}$.
Thus, we consider the spectra $C_{EE}(k)$, $C_{EB}(k)$, and $C_{BB}(k)$.
In some cases, we consider one-point correlators, which are equal to
the integrated spectrum, i.e., $\bra{XY}_{xy}=\int C_{XY}(k)\,\dd k$.
In the following, when we sometimes talk about $EE$ or $BB$ correlations,
we always mean the spectral correlation functions $C_{EE}(k)$ and
$C_{BB}(k)$.

\subsection{Two-scale analysis}
\label{PolarizationAnalysis}

In the Sun, we expect opposite signs of the $EB$ correlation in the
northern and southern hemispheres.
An analogous situation has been encountered previously in connection
with magnetic helicity measurements.
To prevent cancelation from contributions of opposite signs coming
systematically from the two hemispheres, one has to allow for a
corresponding modulation of the sign between the hemispheres.
We refer here to the work of \cite{RS75} for the general formalism in
the context of dynamo theory and to \cite{BPS17} for the application to
observational data similar to those discussed here.
The two-scale formalism has so far only been developed for Cartesian
geometry, but it is conceivable that it can also be extended to spherical
harmonics.

Here we assume that the $x$ direction points in longitude and the
$y$ direction points in latitude.
To account for a sinusoidal modulation in latitude proportional to
$\sin K_y y$, we compute the following generalized spectrum as
\begin{equation}
C_{XY}(\KK,k)=\int_0^{2\pi}\tilde{X}(\kk+\half\KK)\,\tilde{Y}^*(\kk-\half\KK)\,k\,\dd\phi_k,
\label{KKnull}
\end{equation}
and plot $-\Imag \, C_{XY}(\KK,k)$ versus $k$ for $\KK=(0,K_y)$; see
also \cite{Singh18} and \cite{ZB18} for recent applications.

\subsection{A simple example}

\begin{figure}[t!]\begin{center}
\includegraphics[width=\columnwidth]{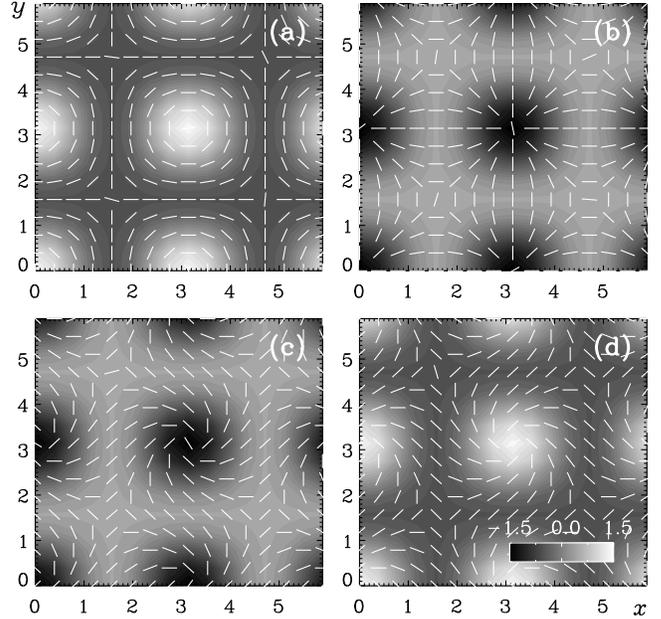}
\end{center}\caption[]{
(a) and (b): pure $E$ polarization for cases
$(f_0,g_0)=(1,0)$ and $(0,1)$, respectively.
(c) and (d): pure B polarization for cases
$(f_0,g_0)=(1,\pm1)$, for upper and lower
signs, respectively.
}\label{p}\end{figure}

We consider gradient- and curl-type vector fields \citep{Dur08}
\begin{equation}
F_i(x,y)=\partial_i f,\quad
G_i(x,y)=\epsilon_{ij}\partial_j g,
\end{equation}
using
\begin{equation}
f=f_0\cos kx\cos ky,\quad
g=g_0\cos kx\cos ky.
\end{equation}
The two-dimensional projection of an otherwise three-dimensional
magnetic field in this model is given by $\bb(x,y)=\FF+\GG$.
Here, $\epsilon_{ij}$ is the totally antisymmetric tensor in two
dimensions, so $\epsilon_{12}=1$, $\epsilon_{21}=-1$, and zero otherwise.
Assuming $k=1$ in a domain $-\pi<(x,y)<\pi$, we have
\begin{equation}
\FF=-f_0\pmatrix{ \sin x\cos y\cr \cos x\sin y},\quad
\GG= g_0\pmatrix{-\cos x\sin y\cr+\sin x\cos y}.
\end{equation}
In \Fig{p} we show examples of polarization maps for {\em different
combinations} of the coefficients $(f_0,g_0)$.
The polarization vectors correspond to $\bb$ vectors without an arrow.
Pure $E$ polarization occurs whenever either $f_0$ or $g_0$ vanish, whereas
pure $B$ polarization occurs whenever $|f_0|=|g_0|$.
Thus, there is no direct correspondence between gradient- and curl-type
vector fields and gradient- and curl-type polarization.
Thus, Equation~(5.84) of \cite{Dur08} is incorrect
(R.~Durrer, private communication).

\begin{figure}[t!]\begin{center}
\includegraphics[width=\columnwidth]{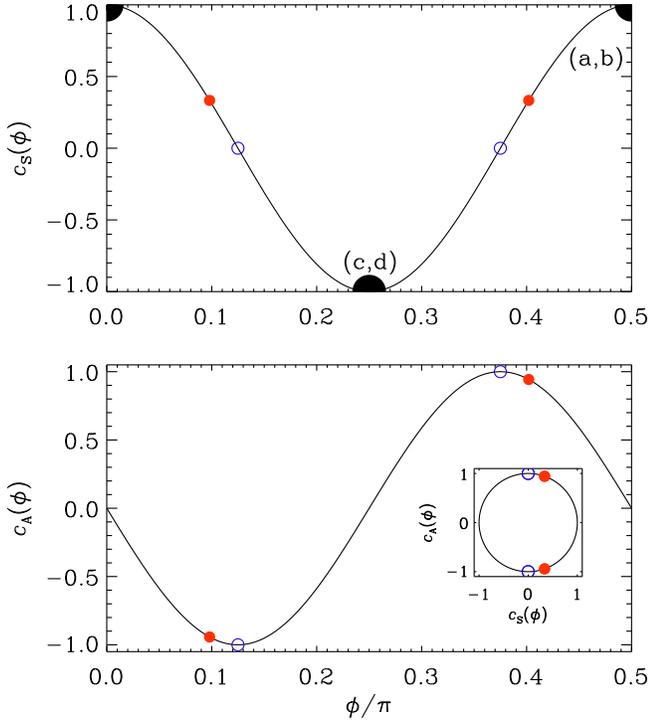}
\end{center}\caption[]{
Dependence of real-space correlations
$c_{\rm S}(\phi)$ (top) and $c_{\rm A}(\phi)$ (bottom) on $\phi$.
The inset shows a parametric representation
of $c_{\rm A}(\phi)$ versus $c_{\rm S}(\phi)$.
The open blue and filled red symbols denote the points where
$\bra{E^2}/\bra{B^2}=1$ and $2$, respectively.
The black filled symbols denote the examples of pure $E$ polarizations
in \Figp{p}{a+b} and pure $B$ polarizations in \Figp{p}{c+d}.
}\label{pscan}\end{figure}

It is convenient to define normalized symmetric and antisymmetric
polarization correlations as
\begin{equation}
c_{\rm S}=\bra{E^2-B^2}\left/\bra{E^2+B^2}\right.
\end{equation}
and
\begin{equation}
c_{\rm A}=2\bra{EB}\left/\bra{E^2+B^2}\right.
\end{equation}
and display them as a function of $\phi$ with
$(f_0,g_0)=(\cos\phi,\sin\phi)$.
Here angle brackets denote averaging over the $xy$ plane.
The resulting polarization maps are shown in \Fig{pscan}.
In this model, the points in a parametric representation of $c_{\rm A}$
versus $c_{\rm S}$ lie on a closed, nearly circular line; see the inset
of \Fig{pscan}.

Pure $E$ polarization implies $c_{\rm S}=1$, while
pure $B$ polarization implies $c_{\rm S}=-1$.
In both cases, we have $c_{\rm A}=0$.
Furthermore, the case $c_{\rm S}=0$
(which coincides with $c_{\rm A}=\pm1$)
corresponds to $\bra{E^2}/\bra{B^2}=1$.
This is what was theoretically expected in the case of dust polarization
as a probe of ISM turbulence; see \cite{CHK17}.
However, of particular interest is now the case $\bra{E^2}/\bra{B^2}=2$,
which has been detected in foreground polarization with {\em Planck}
\citep{Adam16,Akrami18}.
In our model, this implies $c_{\rm S}=1/3$ with $c_{\rm A}=\pm2\sqrt{2}/3$.

Analogously to the real-space coefficients $c_{\rm S}=0$ and
$c_{\rm A}=0$, we define normalized spectra as
\begin{equation}
c_{\rm S}(k)={C_{EE}(k)-C_{BB}(k)\over C_{EE}(k)+C_{BB}(k)}
\end{equation}
and
\begin{equation}
c_{\rm A}(k)={2C_{EB}(k)\over C_{EE}(k)+C_{BB}(k)}.
\end{equation}
Unless noted otherwise $E$ and $B$ have been obtained
from simulations through integration along the $z$ direction; see
\Eq{LOSintegr}.

\begin{figure*}[t!]\begin{center}
\includegraphics[width=\textwidth]{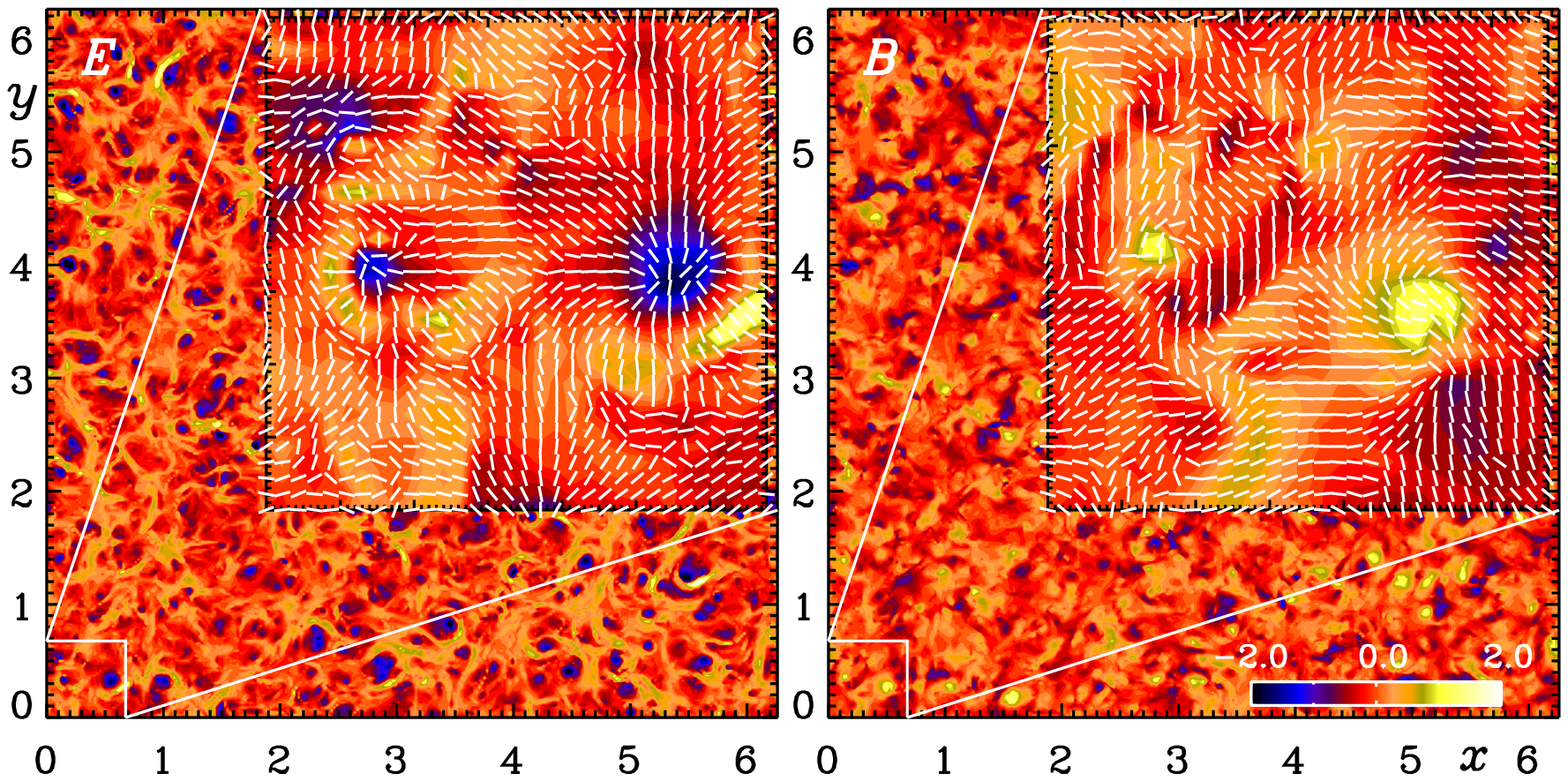}
\end{center}\caption[]{
$E$-mode (left) and $B$-mode (right) polarization for isotropic fully
helical magnetohydrodynamic turbulence using an $xy$ slice of $E$ and $B$
from \cite{BK17} (their Figures~4d--f, for $\Pm=100$).
Dark (light) shades indicate negative (positive) velocity.
In each panel, the insets show an enlarged portion where we also show
the $E$ and $B$ polarization vectors.
}\label{peb4}\end{figure*}

\begin{figure*}[t!]\begin{center}
\includegraphics[width=\textwidth]{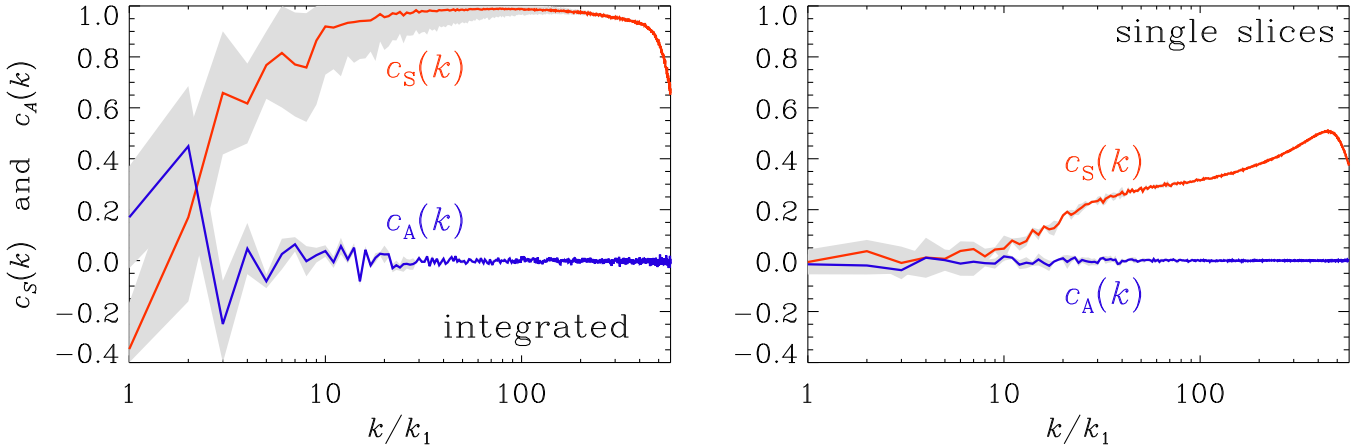}
\end{center}\caption[]{
Spectral correlation functions $c_{\rm S}(k)$ and $c_{\rm A}(k)$ using
line-of-sight integrated polarization (left) and single slice data
(right) computed from decaying isotropic turbulence of \cite{BK17}.
Error margins are indicated in gray.
}\label{pEBmodes_tot}\end{figure*}

\section{Numerical simulations}

\subsection{Isotropic turbulence simulations}
\label{IsotropicTurbulence}

Astrophysical turbulence comes in a multitude of different forms:
it can be helical or nonhelical, it can be magnetically dominated or
subdominant, it can possess cross helicity, with a systematic alignment.
These turbulence simulations provide a means of performing experiments in
a variety of circumstances and environments.
Here we use three-dimensional simulation data of isotropic MHD turbulence
and consider separately all $xy$ planes.
The simulations describe decaying MHD turbulence with magnetic helicity
in the magnetically dominated case.

In the context of early universe turbulence, we have studied decaying
MHD turbulence which is magnetically dominated, i.e., the
magnetic energy exceeds the kinetic energy by typically a factor of ten
\citep{BKMRPTV17}.
The turbulence is then mostly driven by the Lorentz force.
The resulting $E$- and $B$-mode polarizations for individual $xy$ slices
are shown in \Fig{peb4} for a particular example.
We avoid here using forced turbulence, because in the helical case the
magnetic field would become bihelical, i.e., it has opposite signs of
magnetic helicity at different wavenumbers \citep{Bra01}.
Instead, we use decaying hydromagnetic turbulence where with a helical
initial field, the
sign of magnetic helicity is always the same at all wavenumbers
\citep{KBJ11,PB12b}.
This makes the interpretation of the data more straightforward.
In some cases, we also compare with nonhelical initial fields, but the two
turn out to be rather similar with respect to both the $EE/BB$ correlation ratio
and the $EB$ cross correlation.

It is interesting to note that, even though the turbulence is nearly fully
helical with a fractional helicity of about 98\%, the $EB$ correlation,
as quantified by $c_{\rm A}(k)$, is actually zero; see \Fig{pEBmodes_tot}.
This was also confirmed for helical magnetic fields threading interstellar
filamentary structures; see the recent work of \cite{Bracco18}.
In hindsight, and as already discussed in the introduction, this is not
surprising because the parity-odd polarization, as measured by the $EB$
cross correlation characterizes the shape of two-dimensional structures
on a surface---not in a three-dimensional volume.
Thus, it can distinguish between the shapes of the two letters {\sf p}
and {\sf q}, which are mirror images of one another.
In the solar context, one may think of an arrangement of three spots
of different magnetic field strengths on a plane surface.
This arrangement implies a certain sign of magnetic helicity on one
side of the surface, as was recently demonstrated by \cite{BB18}.
In three dimensions, however we can flip any structure and view it from
the backside, provided both directions are physically equivalent, which
is the case when the system is homogeneous.
The superposition of flipped and unflipped versions results in a vanishing
$c_{\rm A}(k)$.

\begin{figure*}[t!]\begin{center}
\includegraphics[width=\textwidth]{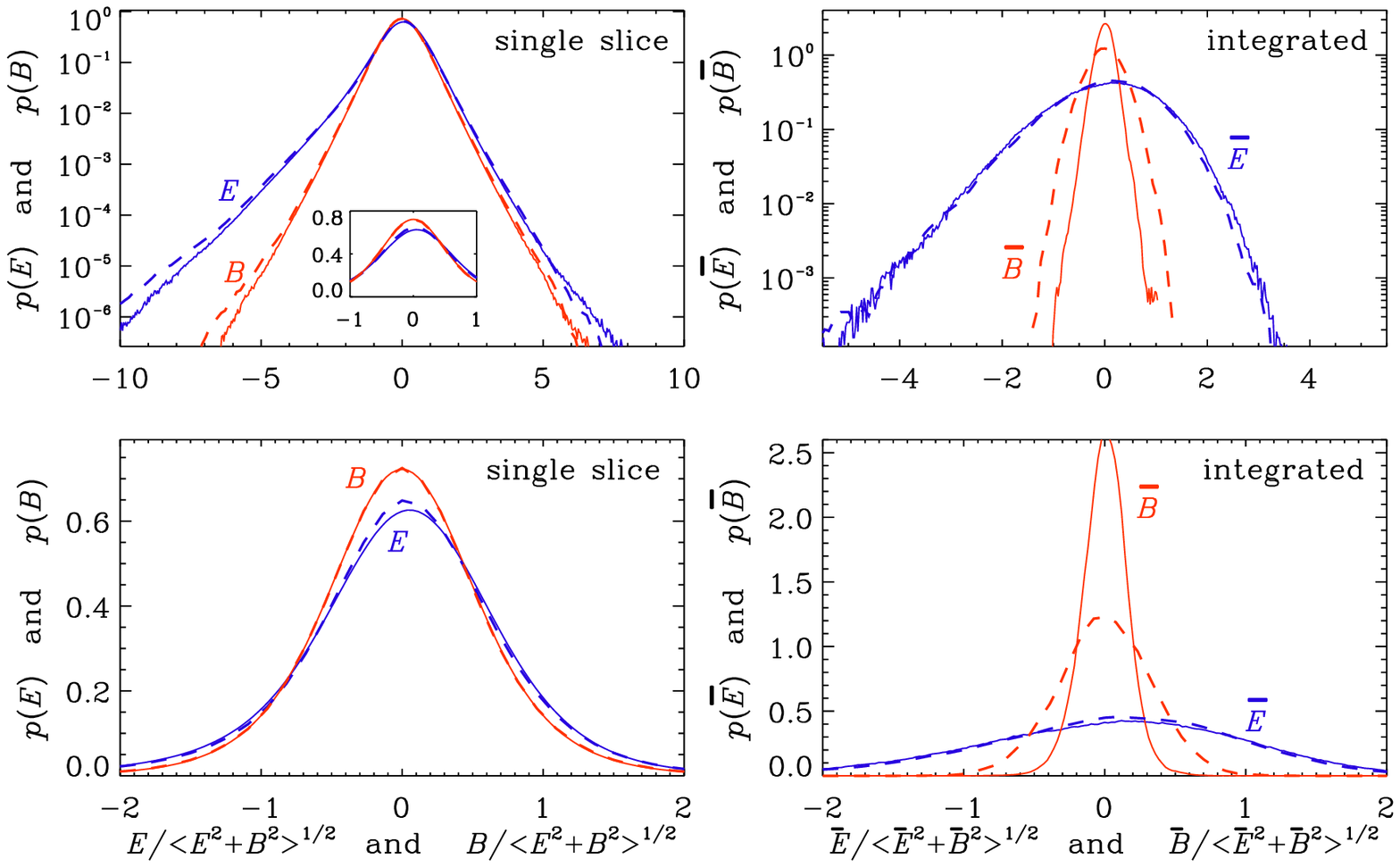}
\end{center}\caption[]{
Probability density functions of $E$ and $B$ polarization (left) and those of
$\meanE$ and $\meanB$ (right) in semilogarithmic (top) and
linear representations (bottom) for helical turbulence
\citep[the runs shown in Figures~4d--f of][solid lines]{BK17} and
nonhelical turbulence \citep[Run~A of][dashed lines]{BKMRPTV17}.
}\label{phisto}\end{figure*}

\begin{figure*}[t!]\begin{center}
\includegraphics[width=\textwidth]{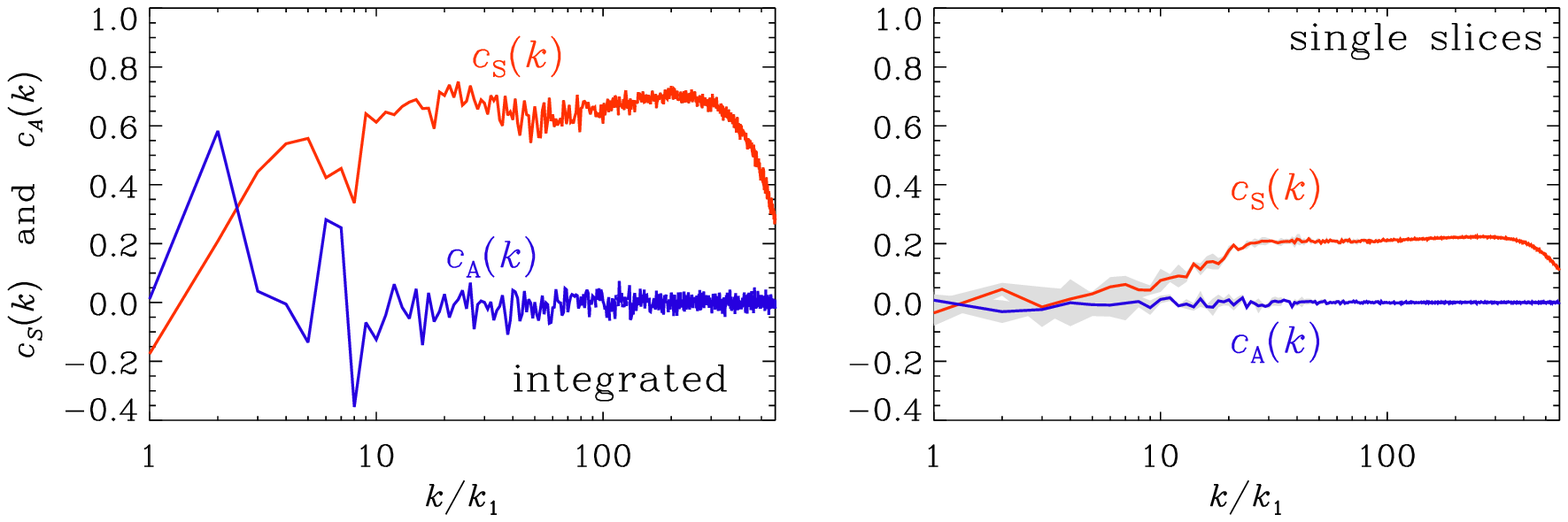}
\end{center}\caption[]{
Similar to \Fig{pEBmodes_tot}, but for dust polarization.
}\label{pEBmodes_tot_dust}\end{figure*}

\begin{figure*}[t!]\begin{center}
\includegraphics[width=\textwidth]{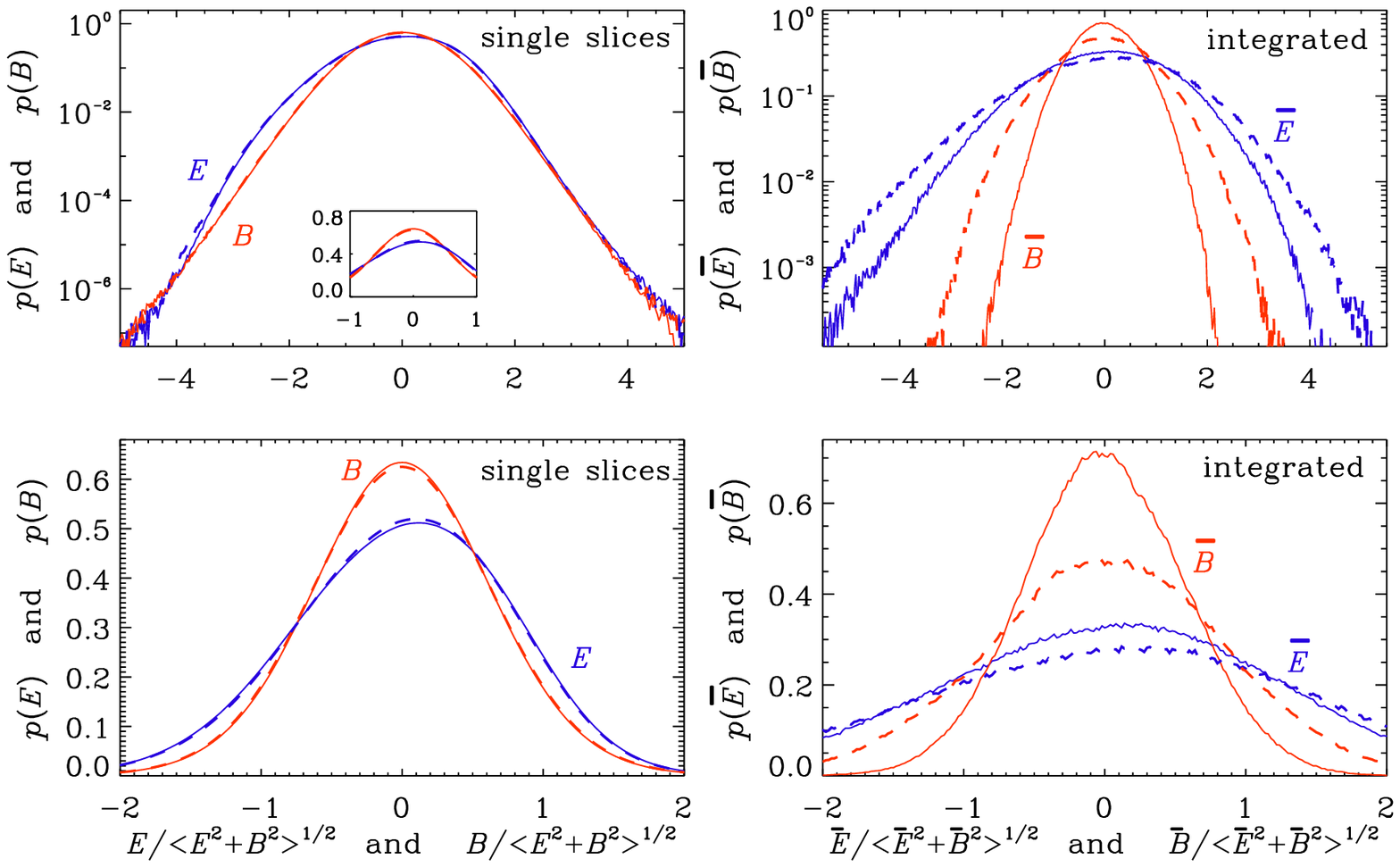}
\end{center}\caption[]{
Similar to \Fig{phisto}, but for dust polarization.
}\label{phisto_dust}\end{figure*}

In \Fig{pEBmodes_tot} we also see that $c_{\rm S}(k)$, based on the
line-of-sight integral in \Eq{LOSintegr}, approaches unity.
In other words, the $EE$ polarization exceeds the $BB$ polarization by
a factor of over a hundred in this case.
This is surprising, because in each of the individual planes, e.g., that
shown in \Fig{peb4}, the $EE$ correlation exceeds the $BB$ correlation
only by a factor of about 2 at $k/k_1\approx30$; see the second panel of
\Fig{pEBmodes_tot}, which was also what was found by \cite{Kritsuk18}
using realistic simulations of supersonic turbulence.
Here and elsewhere, error margins have been computed by using any one third
of the original data and estimate the error as the largest departure from
the full average.

To understand the reason for this, we must look for the possibility of
excessive and preferential cancelation in $B(x,y)$ compared to $E(x,y)$.
In this connection, we recall that, since the transformation from
$(Q,U)$ to $(E,B)$ is a linear one, the line-of-sight integral in
\Eq{LOSintegr} can also be carried out over $E+\ii B$, which is what
we do when we talk about preferential cancelation in $B$ compared to $E$.

\begin{figure}[t!]\begin{center}
\includegraphics[width=\columnwidth]{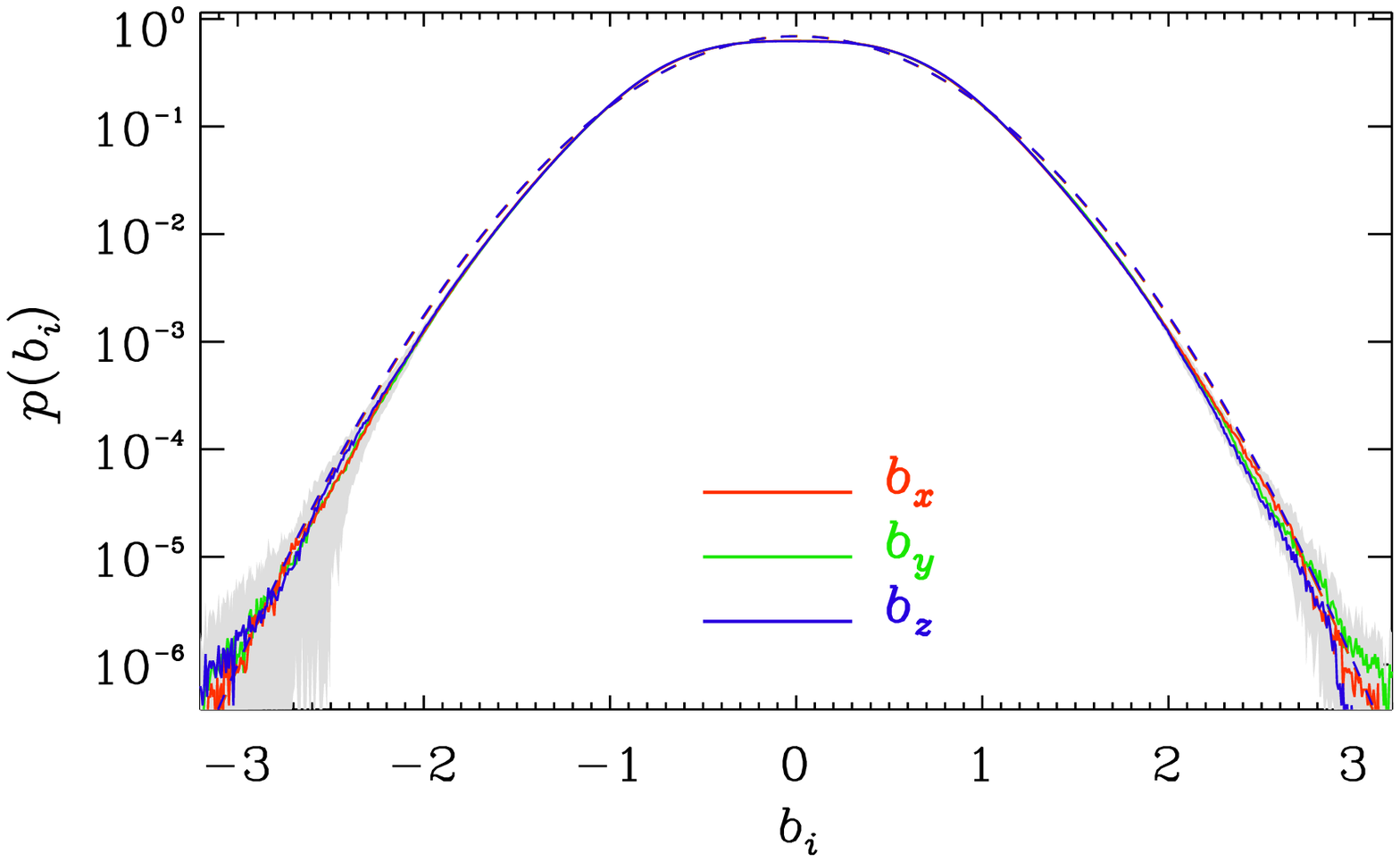}
\end{center}\caption[]{
Histogram of the three components of the magnetic field for the
helical turbulence run.
}\label{phisto_bb}\end{figure}

\begin{figure*}[t!]\begin{center}
\includegraphics[width=.98\textwidth]{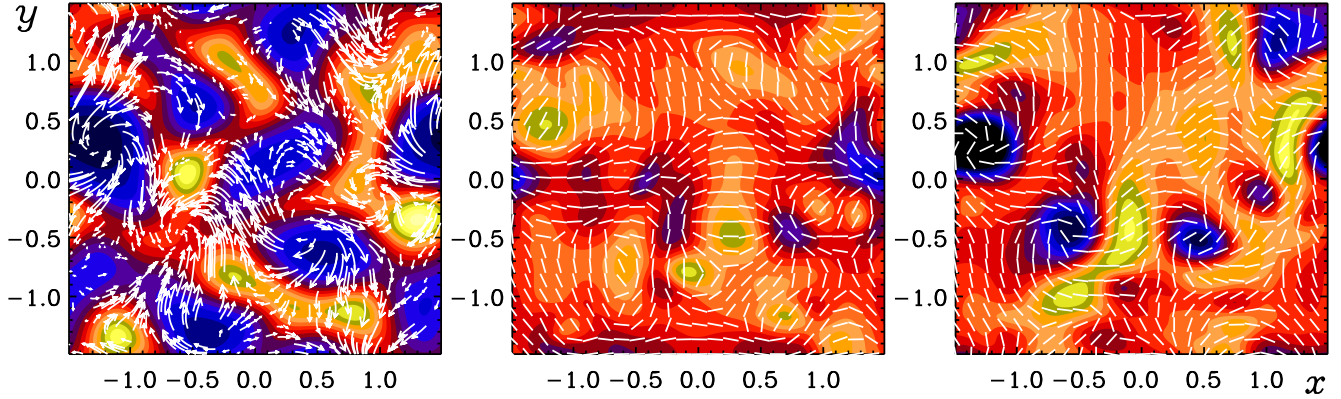}
\end{center}\caption[]{
Magnetic field vectors and line-of-sight component (color-coded; left)
as well as $E$-mode (middle) and $B$-mode polarization (right)
in rotating convection viewed from the top onto the convecting layer.
}\label{pslice}\end{figure*}

\begin{table}[b!]\caption{
Variance, skewness, and kurtosis for the distributions shown
in \Fig{phisto}.
}\vspace{12pt}\centerline{\begin{tabular}{cc|rrrr}
quantity & helical? & $E\;$ & $B\;$ & $\meanE\;$ & $\meanB\;$ \\
\hline
$\sigma$ & yes & $ 0.77$ & $0.63$ & $ 1.51$ & $ 0.25$ \\
         &  no & $ 0.77$ & $0.64$ & $ 1.65$ & $ 0.58$ \\
$\SKEW$  & yes & $-0.55$ & $0.00$ & $-0.45$ & $-0.11$ \\
         &  no & $-0.61$ & $0.00$ & $-0.53$ & $-0.05$ \\
$\KURT$  & yes &   3.09  &  1.54  &   0.56  &   0.58  \\
         &  no &   3.87  &  1.69  &   0.84  &   0.08  \\
\label{Tskew}\end{tabular}}\end{table}

\begin{table}[b!]\caption{
Similar to \Tab{Tskew}, but for the case of dust emission
shown in \Fig{phisto_dust}.
}\vspace{12pt}\centerline{\begin{tabular}{cc|rrrr}
quantity & helical? & $E\;$ & $B\;$ & $\meanE\;$ & $\meanB\;$ \\
\hline
$\sigma$ & yes & $ 0.76$ & $0.65$ & $ 1.21$ & $ 0.56$ \\
         &  no & $ 0.76$ & $0.65$ & $ 1.45$ & $ 0.85$ \\
$\SKEW$  & yes & $-0.19$ & $0.00$ & $-0.19$ & $ 0.01$ \\
         &  no & $-0.20$ & $0.00$ & $-0.24$ & $-0.04$ \\
$\KURT$  & yes & $-0.02$ & $0.26$ & $ 0.07$ & $-0.04$ \\
         &  no & $+0.03$ & $0.24$ & $ 0.16$ & $ 0.00$ \\
\label{Tskew_dust}\end{tabular}}\end{table}

In \Fig{phisto}, we show the probability density functions of $E(x,y,z)$
and $B(x,y,z)$ and compare them with those of the line-of-sight or $z$-integrated
values that we denote here by $\meanE(x,y)$ and $\meanB(x,y)$.
Their variances are $\sigma_E^2=\bra{E^2}-\bra{E}^2$ and
$\sigma_B^2=\bra{B^2}-\bra{B}^2$.
In all cases, the averages are negligible, i.e.,
$\bra{E}^2\approx0$ and $\bra{B}^2\approx0$.
It turns out that, while $B(x,y,z)$ and $\meanB(x,y)$ are symmetric
about zero, $E(x,y,z)$ and $\meanE(x,y)$ are not.
This is quantified by the skewness,
\EQ
\SKEW(E)=\bra{E^3}/\sigma_E^3,\quad
\SKEW(B)=\bra{B^3}/\sigma_B^3.
\EN
These values are listed in \Tab{Tskew} both for helical and
nonhelical turbulence.
These simulations correspond to the runs shown in Figures~4d--f
of \cite{BK17} for the helical case and Run~A of \cite{BKMRPTV17}
for the nonhelical case.
For completeness, we also list there the kurtoses of those fields,
which are defined as
\EQ
\KURT(E)=\bra{E^4}/\sigma_E^4-3,\quad
\KURT(B)=\bra{B^4}/\sigma_B^4-3.
\EN
The consequences of a non-vanishing skewness of $E$ become clear when
looking at the probability density functions of $\meanE(x,y)$ and
$\meanB(x,y)$ in \Fig{phisto}, which show a dramatic difference for
large values where $|E|>\sigma_E$, because now positive and negative
pairs of equal strengths have different abundance or probability and do
not cancel.
The reason for this asymmetry lies in the nature of turbulence, which
has a preference of producing large tails of negative $E$ polarization,
which corresponds to a preference of radial over circular patterns.

In the results presented above, we have assumed that the local emissivity
$\epsilon$ is proportional to $\bb^2$, but this is not realistic in all
astrophysical contexts as for instance in the case of dust polarization,
which is the case for which an enhanced $EE/BB$ correlation
ratio has been found.
In \Fig{pEBmodes_tot_dust}, we show that for constant $\epsilon$, i.e.,
independent of $|\bb|$, we still find $c_{\rm S}>0$, but it is now no
longer so close to unity as in the case when $\epsilon\propto\bb^2$.
Instead, we have $c_{\rm S}\approx0.6$ for intermediate values of $k$,
which corresponds to $C_{EE}/C_{BB}\approx7$.
The result for individual slices is, however, less strongly affected by
the choice of $\epsilon$.

The corresponding probability density functions are shown in
\Fig{phisto_dust}.
We see that the basic asymmetry of the probability density function of $E$
still persists both for individual slices and for the integrated maps, but
the tails of the distribution are now less extended; see \Tab{Tskew_dust}
for the corresponding values of skewness and kurtosis.
As already explained above, the asymmetry in $E$ results here from a
dominance of circular patterns.
However, even a preference of radial patterns would cause asymmetry, albeit
with the other sign.
Any such asymmetry would always lead to an excess of $EE$ correlations
over $BB$ correlations and hence an enhanced $EE/BB$ ratio.

The relative importance of radial patterns over circular ones
is a qualitatively new property of turbulent motions that needs to be
studied further.
It does not imply any asymmetry in the individual components of the magnetic
field, as shown in \Fig{phisto_bb}.

\subsection{Convection}

Next, we perform hydrodynamic simulations with gravity $\grav=(0,0,-g)$
and angular velocity $\OO=(0,0,\Omega)$ in a layer
$z_{\rm bot}\leq z\leq z_{\rm top}$, heated from below.
Here $z_{\rm top}-z_{\rm bot}\equiv d$ is the thickness of the layer.
The governing equations for density $\rho$, velocity $\uu$, the specific
entropy $S$, and the magnetic vector potential $\aaaa$ are given by
\begin{equation}
{\DD\ln\rho\over\DD t}=-\nab\cdot\uu,
\end{equation}
\begin{equation}
\rho{\DD\uu\over\DD t}=-\nab P+\rho\grav-2\OO\times\rho\uu
+\jj\times\bb+\nab\cdot(2\nu\rho\SSSS),
\end{equation}
\begin{equation}
\rho T{\DD S\over\DD t}=K\nabla^2 T+\eta\mu_0\jj^2+2\nu\rho\SSSS^2,
\end{equation}
\begin{equation}
{\partial\aaaa\over\partial t}=\uu\times\bb+\eta\nabla^2\aaaa,
\end{equation}
where $P$ is the pressure with $S=c_v\ln P-c_p\ln\rho$, which is defined
up to some additive constant, $c_p$ and $c_v$ are the specific heats at
constant pressure and density, respectively, $T$ is the temperature with
$P/\rho=(c_p-c_v)T$ being the ideal gas equation of state,
$K$ is the thermal diffusivity, $\nu$ is the kinematic viscosity,
$\eta$ is the magnetic diffusivity, $\bb=\bb_0+\nab\times\aaaa$
is the magnetic field with $\bb_0$ being the imposed field,
$\jj=\nab\times\bb/\mu_0$ is the current density that was already defined
in the introduction.

\begin{table}[b!]\caption{
Parameters for convection simulations.
}\vspace{12pt}\centerline{\begin{tabular}{cccccc}
Run & Ra & $\nu$ & $\lambda$ & $\urms$ & Res. \\
\hline
A &  3600 & 0.01  & 0.35 & 0.050 & $144^2\times48$ \\
B & 14400 & 0.005 & 0.42 & 0.045 & $288^2\times96$ \\
\label{Tconv}\end{tabular}}\end{table}

\begin{figure}[t!]\begin{center}
\includegraphics[width=\columnwidth]{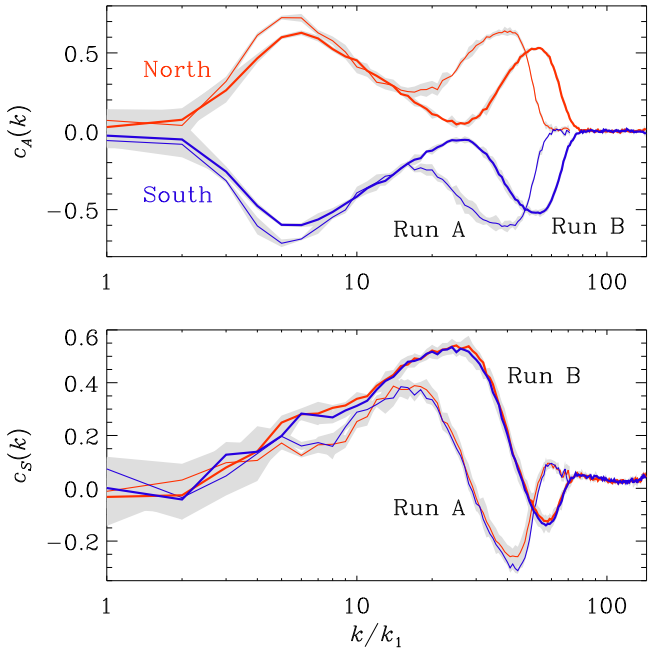}
\end{center}\caption[]{
$c_{\rm A}(k)$ (top) and $c_{\rm S}(k)$ (bottom) for convection
simulations corresponding to the northern (red) and southern (blue)
hemispheres for Runs~A (thin lines) and B (thick lines).
The gray shades indicate error bars obtained from the statistics
over about 50 snapshots covering a time interval of about 1000
time units.
}\label{pvarEBmodes_comp}\end{figure}

We adopt a polytropic stratification with background temperature
$T=-gz/\cp$, so $T=0$ at $z=0$.
We fix $d$ and choose $|z_{\rm top}|/d$ to set the degree of stratification.
The smaller $|z_{\rm top}|/d$, the stronger is the stratification, i.e.,
the stronger is the temperature contrast.
In the following we choose $|z_{\rm top}|/d=0.1$, so the temperature
changes by a factor of 10; see \cite{HTM84} for a similar setup.
We choose $\grav\cdot\OO$ to be either negative or positive,
corresponding to the northern or southern hemispheres, respectively.
A vertical magnetic field is imposed and tangled by this velocity field.
The simulation setup is similar to that of \cite{HT88}, except that they
did not include rotation, which makes our present simulations therefore
closer to those of \cite{BNPST90}, which did include rotation.

In the following, we denote by $\rho_0$ the density at $z=-d$.
Some of the parameters are listed in \Tab{Tconv}.
The imposed magnetic field points in the $z$ direction
and is given by $B_{0z}=0.02\Beq$, where $\Beq^2=\mu_0\rho_0gd$
is the thermal equipartition field strength.
We use $\Omega=0.2\,(g/d)^{1/2}$ in all cases.
The Rayleigh number is defined as
$(gd^4\rho_{\rm m}/\nu K)\,(\dd s/\dd z)_{\rm m}$, where $\rho_{\rm m}$
and $(\dd s/\dd z)_{\rm m}$ are density and the specific entropy of the
hydrostatic solution in the middle of the domain.

Cross-sections of $\bb(x,y)$, $E(x,y)$, and $B(x,y)$ near the surface
are shown in \Fig{pslice} for the results of such a simulation.
One sees cyclonic convection in the northern hemisphere as viewed from
the top, so all converging inflows attain a counterclockwise swirl,
and all diverging outflows are clockwise swirl.
A similar appearance is also attained by the magnetic field.
It would be different when viewing this pattern from beneath that we
would see as a mirror image of the original pattern and therefore the opposite
sign of the $B$ polarization.
The consequence of this can be seen in \Fig{pvarEBmodes_comp}, where we
plot $c_{\rm S}(k)$ and $c_{\rm A}(k)$ for north (red) and south (blue)
for Runs~A and B whose parameters are summarized in \Tab{Tconv}.
There is now a systematic $EB$ correlation, so $c_{\rm A}(k)$ is positive
in the north and negative in the south; \Tab{Tconvhel}.
This is very promising and agrees with our intuition.

\begin{table}[b!]\caption{
Result for convection, as shown in \Fig{pvarEBmodes_comp}.
}\vspace{12pt}\centerline{\begin{tabular}{cc|ccc}
Hemisph.$\!\!$ & $\grav\cdot\OO$ & $c_{\rm A}(k)$ & $\bra{\omega_z u_z}$ & $J_zB_z$ \\
\hline
N & $-$ & $+$ & $-$ & $-$ \\
S & $+$ & $-$ & $+$ & $+$ \\
\label{Tconvhel}\end{tabular}}\end{table}

\begin{figure*}[t!]\begin{center}
\includegraphics[width=\textwidth]{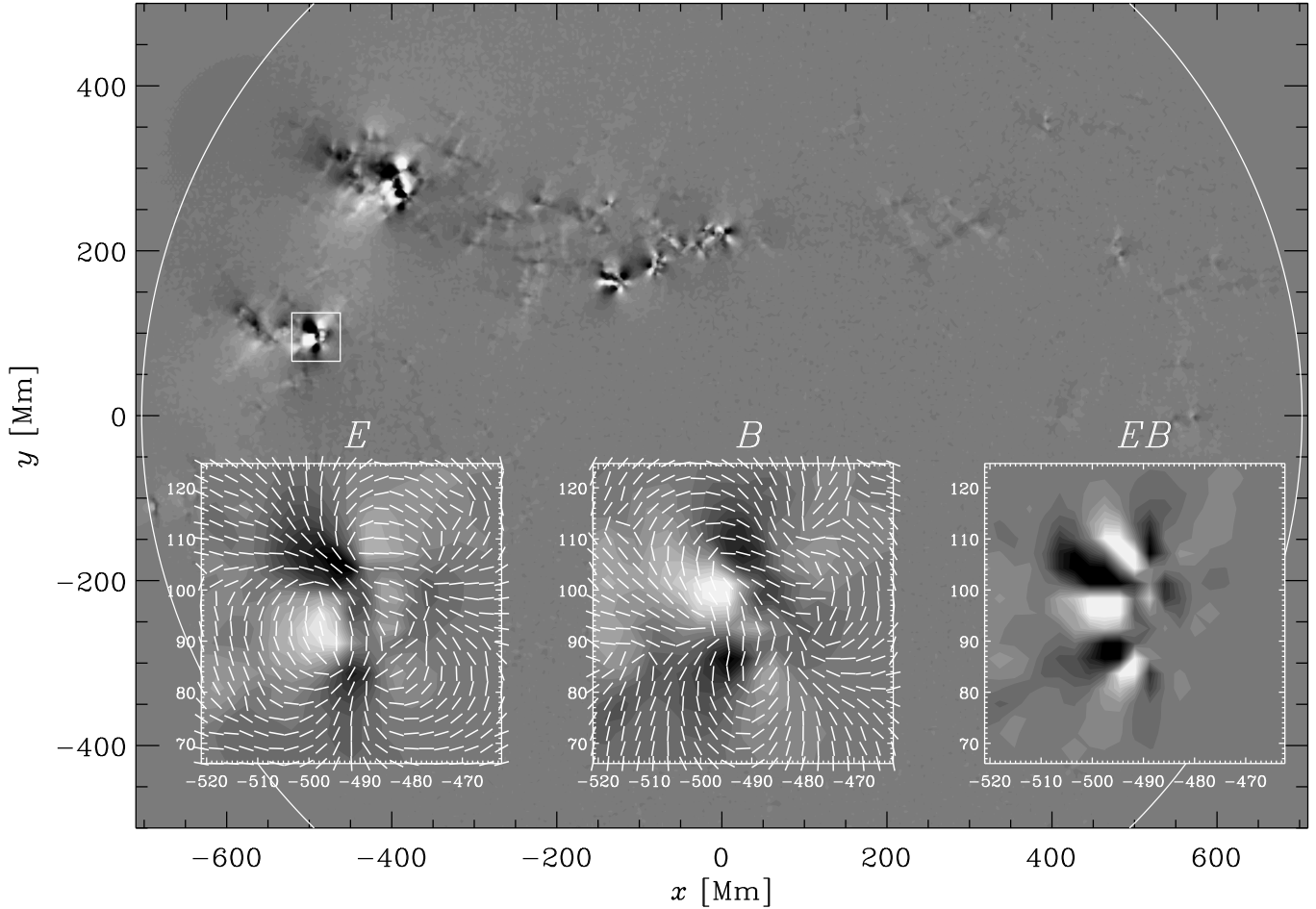}
\end{center}\caption[]{\small{
Solar $E$, $B$, and $EB$ plots in the proximity of AR12325 on 2015--04--16
superimposed on the full disk image of $E$ polarization.
}}\label{p150416_AR12325}\end{figure*}

In rotating convection in the northern hemisphere, we have $\grav\cdot\OO<0$.
Near the upper surface, a downdraft ($u_z<0$) will suffer a counter-clockwise
spin ($\omega_z>0$), so $\omega_z u_z<0$, corresponding to negative
kinetic helicity.
This applies to the sketch shown in \Fig{sketch} (left, for downflows).
Likewise, an updraft ($u_z>0$) will suffer a clockwise
spin ($\omega_z<0$), so again $\omega_z u_z<0$, i.e., the
kinetic helicity is unchanged and its sign equal to that of $\grav\cdot\OO$.
This applies to the sketch shown in \Fig{sketch} (right, for upflows).
Since the polarization vectors have no vector tip, both updrafts and
downdrafts result in the same $E$ and $B$ polarization properties
in each hemisphere.
Therefore $EB$ is positive for $\grav\cdot\OO<0$ (north) and
negative for $\grav\cdot\OO>0$ (south).
In this case, $EB$ does reflect the sign of kinetic helicity,
except that they are opposite to each other.

\section{Prospects of finding solar $EB$ polarization}
\label{SolarData}

We now consider the Stokes $Q$ and $U$ parameters from
the scattering emission on the solar surface.
We ignore Stokes $I$ and $V$ and only look at $Q$ and $U$ at a fixed
wavelength corresponding to the Fe~{\sc I} $630.15\nm$ line
\citep[see][for details of those data]{Hughes}.

\begin{figure*}[t!]\begin{center}
\includegraphics[width=\textwidth]{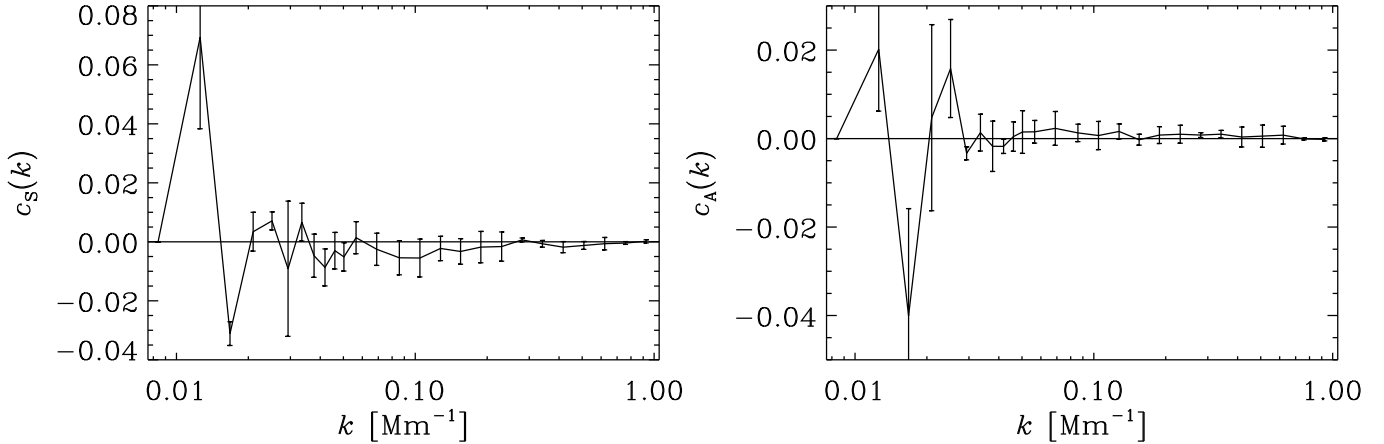}
\end{center}\caption[]{\small{
$c_{\rm S}(k)$ and $c_{\rm A}(k)$ for the time average of
all data from 2010 to 2017.
For wavenumbers above $20\Mm^{-1}$, the data have been averaged
over logarithmically spaced bins.
}}\label{pspecm_all}\end{figure*}

An example is shown in \Fig{p150416_AR12325} using data from the
Synoptic Optical Long-term Investigations of the Sun (SOLIS)
instrument of the NSO Integrated Synoptic Program (NISP).
In the following, we analyze the full disk data such as the one
shown in \Fig{p150416_AR12325}.
In the three insets, we show zoom-ins of $E$, $B$, and the product $EB$
to smaller patch whose location on the solar disk is indicated by a
small square.
In all cases, $E$ and $B$ are computed for the full disk, however.

The resolution of the full disk data is $2048^2$ pixels, but it turned
out that the spectral power at the highest wavenumbers is rather small.
Therefore, we downsampled the data to a resolution of $512^2$ points
and we verified that no essential information is lost in this process.

To have a chance in finding a definite sign, we separate the
signs in the northern and southern hemispheres by using the two-scale
method discussed in Sect.~\ref{PolarizationAnalysis}.
In \Fig{pspecm_all} we show the result for $c_{\rm S}(k)$ and
$c_{\rm A}(k)$ for the years from 2010 to 2017.
The statistical errors are generally large and there are strong sign
changes for $k<0.03\Mm^{-1}$, suggesting that those values are uncertain.
There is a short range of positive values of $c_{\rm A}(k)$ around
$0.04\Mm^{-1}\la k\la0.1\Mm^{-1}$, but those values are still compatible
with zero within error bars.
This somewhat unexpected result remains subject to further investigations.
As seen from \Tab{Tconvhel}, positive values of $c_{\rm A}(k)$
correspond to negative magnetic helicity, which is expected for
the northern hemisphere and compatible with our two-scale analysis,
where the sign corresponds to that of the northern hemisphere.
The wavenumber interval from $0.04$ to $0.1\Mm^{-1}$ agrees with that
where most of the magnetic power has previously been found from the
SOLIS data; see \cite{Singh18}.
It might therefore be useful to target further work to this wavenumber
range.

We also see that $c_{\rm S}(k)$ is fluctuating around zero.
This shows that the $EE/BB$ correlation ratio is about unity, which
is thus quite different from the {\em Planck} results for dust polarization.
This suggests that the effect of line-of-sight integration discussed in
\Sec{IsotropicTurbulence} is here unimportant and could be a consequence
of the optical thickness being large.

\section{Conclusions}

Our work has identified an important factor governing the enhanced ratio
of $EE$ to $BB$ polarization: a strongly asymmetric $E$ distribution
for helical (nonhelical) turbulence with a skewness of $-0.55$ ($-0.61$)
and $-0.19$ ($-0.20$) for $\epsilon\propto\bb^2$ and $\epsilon=\const$,
respectively, compared with an unskewed $B$ distribution.
This implies that, depending on the extent of the line-of-sight
integration, there will be less cancelation of $E$ compared to $B$,
which explains the enhanced  $EE$ to $BB$ ratio.
This was previously explained in terms of Alfv\'en waves in magnetically
dominated flows \citep{KLP17}.

Under inhomogeneous conditions, the $EB$ cross correlation is found to
be a meaningful proxy of kinetic and magnetic helicity.
We have shown that such conditions are found in stratified convection
in the presence of rotation.
This became clear from the sketch shown in \Fig{sketch}.
Homogeneous systems, by contrast, are unable to produce any net $EB$
cross correlation, even if the turbulence is fully helical.
This is because, with respect to a given line of sight,
a helical eddy can face the observer at different viewing angles, where
$B$ can attain positive and negative values, depending on which side of the
plane the observer is facing, while the $E$ polarization can be similar
in both cases, independently of the viewing angle.
For convection, on the other hand, owing to inhomogeneity, it is
impossible to find a local plane whose statistical $EB$ correlations
agrees with one that is flipped, so there can be no cancelation.
This was demonstrated by our numerical experiments, which show a dependence
of the $EB$ correlation on the sign of $\grav\cdot\OO$, and thus on the
kinetic and magnetic helicities.

To assess the prospects of determining parity-odd polarization from
solar scattering emission, we have employed the two-scale analysis
to the oppositely helical contributions from north and south.
Unfortunately, a clear antisymmetric spectral correlation could
not be determined as yet.
Even in the $k$ range between $0.04$ and $0.1\Mm^{-1}$,
where most of the magnetic energy is known to reside in the
SOLIS measurements \citep{Singh18}, the positive values obtained
for $c_{\rm A}(k)$ are compatible with zero.
One reason for this poor hemispheric distinction could be that not all
corrections applied to the final vector spectromagnetograph magnetic
field data are included in the spectral data cubes for Stokes $I$, $Q$,
$U$, and $V$ available from the SOLIS website.
This issue needs to be investigated in future work.

\acknowledgements
We thank the referee for a careful assessment of the paper
and for useful comments.
This work has utilized SOLIS data obtained by the NSO Integrated Synoptic
Program (NISP), managed by the National Solar Observatory, which is
operated by the Association of Universities for Research in Astronomy
(AURA), Inc.\ under a cooperative agreement with the National Science
Foundation.
{\em SDO} is a mission for NASA's Living With a Star program.
We acknowledge partial support from the University of Colorado through
its support of the George Ellery Hale visiting faculty appointment,
the National Science Foundation Astrophysics and Astronomy Grant Program
grants AST1615940 \& AST1615100, and the Swiss NSF SCOPES grant IZ7370-152581.
We acknowledge the allocation of computing resources
provided by the Swedish National Allocations Committee at the Center for
Parallel Computers at the Royal Institute of Technology in Stockholm.
This work utilized the Janus supercomputer, which is supported by the National
Science Foundation (award No.\ CNS-0821794), the University of Colorado
Boulder, the University of Colorado Denver, and the National Center for
Atmospheric Research. The Janus supercomputer is operated by the University of
Colorado Boulder.


\vfill\bigskip\noindent\tiny\begin{verbatim}
$Header: /var/cvs/brandenb/tex/sayan/EBmodes/paper.tex,v 1.104 2018/11/21 17:34:59 brandenb Exp $
\end{verbatim}

\end{document}